\documentclass[aps,prd,amssymb,nofootinbib,superscriptaddress,twocolumn]{revtex4}
\usepackage{graphicx}
\usepackage{amsmath}
\usepackage{amssymb}
\usepackage{amsfonts}
\usepackage{bbm}
\usepackage{amsbsy}
\usepackage{color}
\usepackage{cancel}
 \usepackage{ulem}
\usepackage{multirow}
\usepackage{soul}
\usepackage{marginnote}
\usepackage{tikz}
\usepackage{dsfont}
\usepackage{hyperref}
\newcommand{\ket}[1]{\ensuremath{\left|{#1}\right\rangle}}

\begin{document}

\title{Wigner phase of photonic helicity states in the spacetime of the Earth}

\author{Jan Kohlrus}
\affiliation{School of Mathematical Sciences,
University of Nottingham,
University Park,
Nottingham NG7 2RD,
United Kingdom}

\author{Jorma Louko}
\affiliation{School of Mathematical Sciences,
University of Nottingham,
University Park,
Nottingham NG7 2RD,
United Kingdom}

\author{Ivette Fuentes}
\affiliation{School of Mathematical Sciences,
University of Nottingham,
University Park,
Nottingham NG7 2RD,
United Kingdom}
\affiliation{Faculty of Physics, University of Vienna, Boltzmanngasse 5, 1090 Wien, Austria}

\author{David Edward Bruschi}
\affiliation{Faculty of Physics, University of Vienna, Boltzmanngasse 5, 1090 Wien, Austria}
\affiliation{York Centre for Quantum Technologies, Department of Physics, University of York, Heslington, YO10 5DD York, United Kingdom}

\begin{abstract}
We study relativistic effects on polarised photons that travel in a curved spacetime. As a concrete application, we consider photons in the gravitational field of the Earth, on a closed path that starts at a terrestial laboratory, is reflected at one or more satellites, and finally returns to the laboratory. We find that the photons acquire a non-trivial Wigner phase already when the gravitational field is static, such as the Schwarzschild spacetime, where previous studies have found a trivial Wigner phase for closed photon paths. A gauge-invariant description of this nontrivial Wigner phase remains an open question, to be resolved before the formalism can provide predictions for experiments.
\end{abstract}
\maketitle
%\tableofcontents

%---------------------------------------------------------------------------------------------------------------------------------------%
\section{Introduction}
%---------------------------------------------------------------------------------------------------------------------------------------%
The study of the effects of gravity and relativity on quantum features of physical systems has both foundational and practical significance. On the theoretical side, it can aid our search for a fundamental theory of Nature. On the applied side, it can promote the proposal of new experiments that can push the limits of known science. In particular, quantum systems for advanced space science are promising a revolution in our ability to communicate securely \cite{Villoresi3,Ling_QKD,Pan_QKD,Pan_network}, to distribute quantum computing tasks \cite{Pan_entangl,Pan_telep} and to test cutting-edge physics \cite{satexpreview,precision_interferometry}. Polarised photons are at the core of a large number of these recent quantum experiments and technologies involving satellites \cite{Pan_QKD,Pan_network,Pan_entangl,Pan_telep}. However, the theoretical treatment of quantum polarisation, namely helicity, remains unclear in curved spacetime. The inexorable development of increasingly sophisticated quantum technologies in space calls for a proper theoretical foundation of photonic helicity within the framework of general relativity.

A thorough treatment of light's polarisation rotation in flat and curved spacetime exists. In the case of flat spacetime, we find studies of helicity states of photons and the structure of Wigner's little group \cite{Ca.re,WignerBerry,Bradler}, the influence of the detector's motion on entangled helicity states of light \cite{entangledlight,detectmotion} and a QFT formalism for the Einstein-Podolsky-Rosen (EPR) correlations of polarised photons \cite{REPR}. In curved spacetime \cite{brodutch,brodutch_letter, Kopeikin}, since the first inquiry on photon polarisation \cite{Skrotskii}, many physical cases have been considered among which we find: rotation of polarisation induced by astrophysical \cite{rot_GW2} and cosmological gravitational waves \cite{rot_GW1}, photons experiencing vector perturbations in the metric due to the Cosmic Microwave Background (CMB) \cite{rot_CMB1} and quasars \cite{rot_CMB2}, gravitational lensing by massive objects passing through the light's trajectory \cite{rot_GL5,rot_GL8} with relativistic velocities \cite{rot_GL1,rot_GL3}, or with rotation \cite{rot_GL2,rot_GL4,rot_GL6, rot_GL7}. Furthermore, there have been studies of optical interferometry experiments in a near-Earth environment \cite{PPN}, namely with polarisation vectors of two light rays that are compared when they recombine after having travelled through two paths at different altitudes, and therefore under different gravitational potentials.

In this work we look at the effect encoded in the rotation of the polarisation of a pulse of light after it has propagated through a curved background. We employ a general formalism developed in \cite{Pa.Ta.We, Pa.th} and study the evolution of the helicity quantum states of photons sent from Earth to different satellites and finally reflected back to Earth. The helicity states are affected by propagation in a curved spacetime. Surprisingly, we find that the states acquire a non-trivial Wigner phase already in Schwarzschild spacetime, where previous studies found a trivial phase for photon paths that are closed without reflections at satellites [15, 16]. We have however not found a gauge-independent description of these nontrivial Wigner phases, and we regard the formalism incomplete, and inadequate for providing predictions to be compared with experiments, until a gauge-independent formalism is found. Finding such a formalism remains a significant task for future work.

The work is organized as follows: in Section \ref{formalism}, we provide all necessary mathematical tools. In Section \ref{polarisationrotation} we compute explicitly the Wigner rotation for the propagating light pulses. In Section \ref{schemes}, we present two specific operational schemes and we find the Wigner phase picked up by the photons' helicity states along the paths. We conclude on our results in Section \ref{conclusion}.

%---------------------------------------------------------------------------------------------------------------------------------------%
\section{Mathematical formalism}\label{formalism}
%---------------------------------------------------------------------------------------------------------------------------------------%
In this section we introduce the mathematical tools necessary to our work. This includes background spacetime surrounding the Earth, the static observers and their reference frames, the propagation and the polarisation of light rays, and finally the quantum states of the polarised photons. 

Throughout this paper we use natural units $c=G=1$.

%---------------------------------------------------------------------------------------------------------------------------------------%
\subsection{Schwarzschild spacetime and static observers}\label{spacetime}
%---------------------------------------------------------------------------------------------------------------------------------------%

%---------------------------------------------------------------------------------------------------------------------------------------%
\subsubsection{The Schwarzschild spacetime}
%---------------------------------------------------------------------------------------------------------------------------------------%
The spacetime considered in this work is Schwarzschild spacetime, which is the spacetime background for the vacuum surrounding a static spherical mass distribution. This is a good approximation for the Earth where, for the purposes of this work, we can safely ignore its angular momentum and asphericity.  More refined metrics, such as the Kerr \cite{Kerr} or the Hartle-Thorne \cite{Hartle-Thorne} metric, can be employed to take into account these features. We leave more detailed and realistic calculations for further work.

The metric is a symmetric, bilinear form with matrix representation in the usual Schwarzschild coordinates $(t,r,\theta,\phi)$
%\in \mathbb{R} \times  \mathbb{R}_+ \times [0,\pi] \times [0,2\pi[$ 
that reads
\begin{align}\label{metric}
\boldsymbol{g}=\text{diag}\left(-f(r),\frac{1}{f(r)},r^2,r^2 \sin^2{\theta}\right),
\end{align}
where $f(r):=1-\frac{2M}{r}$. The surface of the Earth is placed at $r=R_E$. The extremal values of the polar angle, $\theta=0$ and $\theta=\pi$, indicate the North and South poles respectively, while the equatorial plane is given by $\theta=\pi/2$.

The metric \eqref{metric} is fully parametrised by the Earth's mass $M$. With units restored, we see that the quotient $M/r$ reads $GM/rc^2\ll1$ for any radius $r \geq R_E$, since the Schwarzschild radius $r_S=2M$ has the value $r_S\sim9$mm while $R_E\sim6341$km. In this work we restrict to realistic scenarios where $r \geq R_E$.

%---------------------------------------------------------------------------------------------------------------------------------------%
\subsubsection{Static observers in Schwarzschild spacetime}
%---------------------------------------------------------------------------------------------------------------------------------------%
The observer that will prepare and measure the states in our proposed scheme is a static observer located on the surface of the (non-rotating) Earth. In  Schwarzschild spacetime, there are static observers at any spacetime point with $r>2M$, and their 4-velocity is
\begin{align}\label{static}
\boldsymbol{v_E} =&\, \frac{1}{\sqrt{1-\frac{2M}{r}}} \, \partial_t.
\end{align}
We will consider an observer emitting and measuring polarised quantum signals who is located at a \textit{fixed} radius $r=R_E$. We will need to introduce a family of fiducial static observers all along the light's trajectory in order to track the evolution of the observer's reference frame (i.e., to parallel transport it). This is because the frame in which the polarisation is initially defined by the observer at $r=R_E$ changes as it is carried by the light propagating in the curved spacetime around the Earth. When the polarisation is measured at the end of the light's propagation, the reference frame has changed in general, even if the light is measured at the same location and with the same direction as it was initially emitted.

%---------------------------------------------------------------------------------------------------------------------------------------%
\subsection{Reference frames and tetrads}\label{tetrad_formalism}
%---------------------------------------------------------------------------------------------------------------------------------------%
An observer defines its reference frame mathematically by a tetrad $\boldsymbol{e}$, with tetrad elements $e_{\hat{a}}^{\, \, \, \mu}$ \cite{feliceclarke}. The tetrad relates physical quantities from their expression in the global coordinates of the spacetime  $(t,r,\theta,\phi)$ to the expression of these quantities in the local Cartesian frame of the observer. In other words, in its local frame the observer witnesses a flat, Minkowski spacetime with metric $\boldsymbol{\eta}=\text{diag}(-1,1,1,1)$. The domain of validity of this frame is therefore the spacetime in the vicinity of the observer's world-line, where the effects of curvature can be neglected. In this local frame, the polarisation 4-vector $\boldsymbol{\psi}$ of a light ray passing by will be given components $\psi^{\hat{a}}=\psi^{\mu}e^{\hat{a}}_{\, \, \, \mu}=\psi^{\mu}e_{\hat{b}}^{\, \, \, \nu} \eta^{\hat{a}\hat{b}} g_{\mu\nu}$.\footnote{Hatted indices denote that the quantity is viewed in the observer's local frame, while quantities with Greek indices are the usual vectors in the global spacetime coordinates basis.} 

As discussed above, we need to keep track of the evolution of the static observer's frame along the path followed by the light rays, in order to be able to properly compare the change in the polarisation of the field. Therefore, we will define a tetrad field $\boldsymbol{e}(\lambda)$ along the light's null geodesic parametrised by the affine parameter $\lambda$.
The tetrad components are obtained explicitly as follows: 
\begin{itemize}
	\item[i)] The timelike component $\boldsymbol{e}_{\hat{0}}$ is the velocity of the observer, in our case $\boldsymbol{e}_{\hat{0}}:=\boldsymbol{v_E}$ defined in \eqref{static};
	\item[ii)] The triad of spacelike components $\boldsymbol{e}_{\hat{1}}$, $\boldsymbol{e}_{\hat{2}}$, $\boldsymbol{e}_{\hat{3}}$ of the tetrad is obtained through the orthonormalisation relations $e_{\hat{a}}^{\, \, \, \mu} e_{\hat{b}}^{\, \, \, \nu}g_{\mu\nu}=\eta_{\hat{a}\hat{b}}$;
	\item[iii)] Each of the three triad components are vectors that have four elements themselves, and these twelve elements are constrained by only nine orthonormalisation relations. Therefore, there is some gauge freedom in the choice of the triad, and thus of the tetrad. This mathematical gauge freedom represents the observer's freedom of choice of the reference frame. In what follows, we will adapt our observer's tetrad to the light's null vector, which will simplify the polarisation rotation computations. To do so we will choose to set the third component of the triad $\boldsymbol{e}_{\hat{3}}$ in a specific way that depends on the trajectory of the light ray. This procedure will be explicitly described in the next subsection. After having adapted the tetrad to the null vector of the emitted light ray, there are then only eight tetrad elements $e_{\hat{a}}^{\, \, \, \mu}$ left to be determined and seven orthonormalisation equations. This implies that all of the remaining gauge freedom is encoded in one single tetrad element, and a choice of two signs due to the second order nature of the two remaining normalisation relations.
\end{itemize}
By extending this procedure to the whole null geodesic followed by the light ray, we obtain the tetrad field for our static observers.

%---------------------------------------------------------------------------------------------------------------------------------------%
\subsection{The light's null vector}
%---------------------------------------------------------------------------------------------------------------------------------------%
In this work we neglect the deviation from the null geodesic due to the photon's helicities, since such corrections are of order $\tilde{\lambda}/b \ll 1$, see \cite{helicity_deviation}, where $\tilde{\lambda}$ is the photon's wavelength and $b$ is the shortest distance between the photon and the center of the Earth, i.e. $b=R_E$. 

%---------------------------------------------------------------------------------------------------------------------------------------%
\subsubsection{General null vector}
%---------------------------------------------------------------------------------------------------------------------------------------%
A light ray defines the path followed by pulses of light and is defined formally by the null vector $\boldsymbol{k}$ tangent to its null geodesic \cite{felicebini}. In the spacetime considered here, such a null vector has the general expression
\begin{align}\label{null}
\boldsymbol{k} =&\, E_p \Bigg( \frac{1}{f(r)} \, \partial_t + \epsilon_r \sqrt{1-f(r)\frac{l_{\phi}^2+\kappa}{r^2}} \, \partial_r \nonumber \\
&\,+ \frac{\epsilon_{\theta}}{r} \sqrt{\frac{\kappa-l_{\phi}^2 \cot^2\theta}{r^2}} \, \partial_{\theta} + \frac{l_{\phi}}{r^2 \sin^2\theta} \, \partial_{\phi} \Bigg),
\end{align}
where $\epsilon_r=\pm1$ depending on whether the light ray ascends or descends the gravitational field of the Earth, and $\epsilon_{\theta}=\pm1$ when the photon's polar angle $\theta$ increases or decreases along its path, respectively.

There are three constants of motion in \eqref{null}. One comes from the stationarity of the metric \eqref{metric}, while the other two are due to the spherical symmetry of the spacetime. The first constant is $E_p$, which gives the value of the energy of the photon as it would be measured by an inertial observer at spacelike infinity. We then have the rescaled azimuthal angular momentum and ``Carter'' constants, $l_{\phi}=L_{\phi}/E_p$ and $\kappa=K/E_p^2$ respectively, that are independent on the energy of the photon encoded in $E_p$. The constants $L_{\phi}$ and $K \geq 0$ are the azimuthal angular momentum and a quantity related to the square of the total angular momentum \cite{Carter} of the photon as seen by an inertial observer at spatial infinity, respectively. The constant of motion $E_p>0$ can easily be related to the photon's frequency $\Omega$ measured by an observer with velocity $\boldsymbol{v}$ at any spacetime point through the relation $\hbar \, \Omega = - \boldsymbol{k} \cdot \boldsymbol{v}$. For example, our static observer \eqref{static} on Earth would measure $E_p = \hbar\,\Omega_E \sqrt{1-\frac{2M}{R_E}}$, where $\Omega_E$ is the photon's frequency measured by the static observer on the surface of the Earth.

It is more complicated to obtain such a simple analytic formula for the two other constants of motion $l_{\phi}$ and $\kappa$. We note that in the special case of radially propagating light rays, i.e. for constant polar angle $\theta$ and longitude angle $\phi$ along the null geodesic, or for an equatorial $\theta=\pi/2$ trajectory, we simply have $\kappa=0$.

%---------------------------------------------------------------------------------------------------------------------------------------%
\subsubsection{Null vector with constant longitude $\phi$}
%---------------------------------------------------------------------------------------------------------------------------------------%

In order to simplify the computations and in particular to provide simple analytical formulas for the Wigner phases, without loss of generality we will assume in Sections \ref{polarisationrotation} and \ref{schemes} that the light rays can be sent with a constant azimuthal angle $\phi$, namely that $l_{\phi}=0$ in \eqref{null}.
For $l_{\phi}=0$, the null vector $\boldsymbol{k}$ takes the simpler expression
{\small
\begin{align}\label{null_constphi}
\boldsymbol{k}=&\, E_p \Bigg(\frac{1}{f(r)} \, \partial_t + \epsilon_r \sqrt{1-f(r)\frac{\kappa}{r^2}} \, \partial_r  + \frac{\epsilon_{\theta} }{r} \sqrt{\frac{\kappa}{r^2}} \, \partial_{\theta} \Bigg).
\end{align}
}Since the light rays are confined to $\phi=$ const. planes, the quantity $\kappa$ now plays the role of the square of the polar angular momentum (i.e., it is related to the $\theta$ polar coordinate). Formally, the constant $\epsilon_{\theta} \sqrt{\kappa}$ is here equivalent to the $l_{\phi}$ constant of motion of a null vector \eqref{null} confined to the equatorial plane $\theta=\pi/2$. To obtain $\kappa$ in this case amounts to finding the single angular-momentum-like constant on a geodesic whose plane is already known. Note that due to the spacetime's spherical symmetry, we could have worked equivalently in any other plane containing the center of the Earth. We chose to work in a $\phi=$ const. plane for simplicity and because the computations for this configuration are easier to extend to the case of the rotating Earth, since most of the rotation effects would be orthogonal to this plane.

%---------------------------------------------------------------------------------------------------------------------------------------%
\subsection{Adapted frames and polarisation rotation in curved spacetime}
%---------------------------------------------------------------------------------------------------------------------------------------%
We have defined the observer's frames and the null vector of light. We now proceed to implement the formalism required to describe the light's polarisation and its rotation in a curved spacetime background.

%---------------------------------------------------------------------------------------------------------------------------------------%
\subsubsection{Polarisation vectors and adapted frames}
%---------------------------------------------------------------------------------------------------------------------------------------%
We have defined the light ray through the null vector \eqref{null}, and we will now define the light's polarisation vector in the standard way \cite{Pa.Ta.We}. The polarisation vector $\boldsymbol{\psi}$ is a spacelike vector orthogonal to the light's null vector $\boldsymbol{k}$, namely $\psi^{\mu} k_{\mu} = \psi^{\hat{a}} k_{\hat{a}} = 0$. Because $\boldsymbol{k}$ is null, i.e., $k^{\mu}k_{\mu}=0$, there exist gauge transformations of the polarisation vector that leave the orthogonality relation invariant. These read
\begin{align}\label{gauge}
\boldsymbol{\psi} \to \boldsymbol{\psi} + C \boldsymbol{k},
\end{align}
where $C$ is an arbitrary real constant. The light's polarisation vector is therefore not uniquely defined by its orthogonality with the light's null vector. This particular gauge freedom is unwelcome as it prevents a unique definition of the polarisation vector. Here we discuss how to eliminate this freedom. We start by noting that we can construct the observer's tetrad in such a way that two of the polarisation components measured in that particular frame would be invariant under gauge transformations like \eqref{gauge}. This construction amounts to adapting the frame to the null vector of the light ray considered. To adapt our tetrad to the light ray, we need to make the simple choice \cite{Pa.Ta.We}
\begin{align}\label{adapting}
\boldsymbol{e}_{\hat{3}}=\frac{\boldsymbol{k}}{k^{\hat{0}}}-\boldsymbol{e}_{\hat{0}}.
\end{align}
Physically, this choice amounts to aligning the third component $\boldsymbol{e}_{\hat{3}}$ of our observer's tetrad with the light's propagation direction. In mathematical terms, that means that we have $k^{\hat{1}}=k^{\hat{2}}=0$ and $k^{\hat{3}}=-k^{\hat{0}}$ in the adapted frame. With this choice for the third component of the triad, the polarisation components in the observer's adapted frame $\psi^{\hat{1}}$ and $\psi^{\hat{2}}$ are gauge invariant under transformations \eqref{gauge}.

%---------------------------------------------------------------------------------------------------------------------------------------%
\subsubsection{Polarisation rotation in curved spacetime}
%---------------------------------------------------------------------------------------------------------------------------------------%
In the adapted frames that we have just defined, the polarisation rotation equations take a simple form \cite{Pa.Ta.We}:
\begin{align}\label{evolution}
\frac{d\psi^{\hat{A}}}{d\lambda} + u^{\mu} w_{\mu \, \, \, \hat{B}}^{\, \, \, \hat{A}} \psi^{\hat{B}}= 0.
\end{align}
Capital letters $A,B\in\{1, 2\}$ denote the indices of the polarisation's gauge free components. We have defined the rescaled null vector $\boldsymbol{u}=\boldsymbol{k}/k^{\hat{0}}$ and we recall that $\lambda$ is the affine parameter along the light's null geodesic. In these two coupled first order differential equations, we need to compute the Wigner rotation $u^{\mu} w_{\mu \, \, \, \hat{B}}^{\, \, \, \hat{A}}$ in order to obtain the rotation of the polarisation. This term is expressed through to the spin-1 connection as
\begin{align}\label{connection}
w_{\mu \, \, \, \hat{B}}^{\, \, \, \hat{A}}=e^{\hat{A}}_{\, \, \, \rho}\partial_{\mu}e_{\hat{B}}^{\, \, \, \rho} + \Gamma^{\sigma}_{\ \mu \rho} e^{\hat{A}}_{\, \, \, \sigma} e_{\hat{B}}^{\, \, \, \rho}.
\end{align}
In order to compute the Wigner rotation we will thus need to use the expression of the light's null vector \eqref{null}, the adapted tetrad field $\boldsymbol{e}(\lambda)$ of static observers along the light's null geodesic, and the Christoffel symbols defined by the metric components and their derivatives as $\Gamma^{\sigma}_{\ \mu \rho}= g^{\sigma \nu}/2\,(\partial_{\rho} g_{\nu \mu} + \partial_{\mu} g_{\nu \rho} - \partial_{\nu} g_{\mu \rho})$. The explicit expressions for the Christoffel symbols in Schwarzschild spacetime are known \cite{catalogue}. We can simplify \eqref{evolution} by noticing that $\boldsymbol{w}_{\hat{A} \hat{B}}$ is an antisymmetric tensor, and we obtain
\begin{align}\label{evolution2}
\frac{d\psi^{\hat{1}}}{d\lambda} + \tilde{\Psi} \psi^{\hat{2}}= 0, \quad \frac{d\psi^{\hat{2}}}{d\lambda} - \tilde{\Psi} \psi^{\hat{1}}= 0,
\end{align}
with the Wigner rotation $\tilde{\Psi} = u^{\mu} w_{\mu \, \, \, \hat{2}}^{\, \, \, \hat{1}}=-u^{\mu} w_{\mu \, \, \, \hat{1}}^{\, \, \, \hat{2}}$. Notice that the system of coupled equations above can be decoupled by taking derivatives with respect to the affine parameter $\lambda$ and then feeding back the first order equations in to the second order ones. We do not follow this path, but it can be of interest perhaps when employing numerical simulations.

In the next subsection, we will describe how the Wigner rotation of the classical polarisation vector described by the coupled equations \eqref{evolution2} yields a phase change in the quantum state of the photons that propagate through curved spacetime.

%---------------------------------------------------------------------------------------------------------------------------------------%
\subsection{Photonic wave packets and Wigner phases}\label{quantum_formalism}
%---------------------------------------------------------------------------------------------------------------------------------------%
A pulse of light, whether composed by a single photon or many, can be modelled by a wave packet, i.e., a continuous superposition of momentum states weighted by an appropriate shape function. For each of these momentum states, the corresponding helicity state is a superposition of positive and negative helicity eigenstates giving the momentum-helicity eigenstate $\ket{\boldsymbol{p},s}$. If an observer prepares a photon with an equal proportion of positive and negative helicity eigenstates for each momentum state, the initial pure quantum state $\ket{\boldsymbol{\gamma}}$ can be written as
\begin{align}\label{helicity:superposition}
\ket{\boldsymbol{\gamma}}=\frac{1}{\sqrt{2}}\sum\limits_{s=\pm1} \int d\boldsymbol{p} \, \mathrm{F}(\boldsymbol{p}) \ket{\boldsymbol{p},s},
\end{align}
where $\boldsymbol{p}=(k^{\hat{0}}, k^{\hat{1}}, k^{\hat{2}}, k^{\hat{3}})$ is the photon's null vector as seen by the observer in its chosen reference frame, $s$ is the helicity which can only take values $\pm1$ for photons and the function $\mathrm{F}(\boldsymbol{p})=F(\boldsymbol{\vec{p}}) \, \delta(|\boldsymbol{\vec{p}}|^2 -(k^{\hat{0}})^2) \, \bar{\theta}(k^{\hat{0}})$ is the distribution of whole 4-momenta, where $\bar{\theta}$ the step function and $F$ is the distribution function of the 3-momenta. The distribution $F(\boldsymbol{\vec{p}})$ is normalised through $\int d\boldsymbol{\vec{p}} \,\, |F(\boldsymbol{\vec{p}})|^2 = 1$. This implies that the corresponding annihilation operators $\hat{a}_{\vec{p},s}$ of the photons have the usual commutation relations $[\hat{a}_{\vec{p}_1,s_1},\hat{a}_{\vec{p}_2,s_2}^{\dagger}]=\delta(\vec{p}_1-\vec{p}_2)\delta_{s_1s_2}$.

Now, let's assume that the photon with state \eqref{helicity:superposition} has been prepared by the source at $R_E$ and $\boldsymbol{p}$ thus represents the photon's momentum as seen by the observer located at the emission event, parametrised by $\lambda_e$ on the null geodesic. We are interested in knowing the expression for the quantum state of the photon that will be observed by the receiver at affine parameter $\lambda_o$. We will give here a simple procedure to obtain this final state and all details can be found in \cite{Pa.th}. 

Let the photons take a certain path in spacetime, that is the union of propagation along geodesics connected by, for example, mirror reflections on satellites. Let $\hat{\boldsymbol{U}}$ be the operator that encodes the propagation of the photons from the beginning to the end of a specific geodesic segment of the path, denoted by affine parameters $\lambda_1$ and $\lambda_2$ respectively, with here $\lambda_1 = \lambda_e$ and in general $\lambda_e \leq \lambda_1 < \lambda_2 \leq \lambda_o$. 
Each momentum-helicity eigenstate $\ket{\boldsymbol{p},s}$ that constitutes the photon acquires a different Wigner phase factor because of its dependence on both the photon's propagation direction $\boldsymbol{\vec{n}}=\boldsymbol{\vec{p}}/k^{\hat{0}}$ and helicity $s$, see \cite{Pa.Ta.We}. Therefore we have
\begin{align}\label{phase factor}
\hat{\boldsymbol{U}} \ket{\boldsymbol{p},s}=e^{i s \Psi(\boldsymbol{\vec{n}})} \ket{\boldsymbol{p'}, s},
\end{align}
where the prime denotes evaluation at the endpoint of the null geodesic considered. Since every momentum-helicity eigenstate picks up a different phase factor, the quantum state acquires a measurable, not-global phase. In the evolution equation \eqref{phase factor} we find the Wigner phase $\Psi(\boldsymbol{\vec{n}})$ that is obtained by integration of the Wigner rotation along the photon's null geodesic, namely
\begin{align}\label{Wigner phase}
\Psi= \int_{\lambda_1}^{\lambda_2} \tilde{\Psi} \, d\lambda.
\end{align}
We apply the propagator $\hat{\boldsymbol{U}}$ to the initial state \eqref{helicity:superposition} and we use equation \eqref{Wigner phase} to obtain the transformed state
\begin{align}\label{helicity superposition2}
\ket{\boldsymbol{\gamma '}}=\frac{1}{\sqrt{2}} \sum\limits_{s=\pm1} \int d\boldsymbol{p} \, e^{i s \Psi(\boldsymbol{\vec{n}})} \mathrm{F}(\boldsymbol{p}) \ket{\boldsymbol{p'}(\boldsymbol{p}), s}.
\end{align}
We note that the photons have followed a single null geodesic between the points labelled by the affine parameters $\lambda_1$ and $\lambda_2$. This means that the null vectors $\boldsymbol{p'}$ and $\boldsymbol{p}$ are constructed through the same constants of motion, although they are obtained by  evaluation at different spacetime coordinates. This implies that $\boldsymbol{p'}$ is a function of the initial momentum $\boldsymbol{p}$.

We also note that our initial state \eqref{helicity:superposition} prepared at $\lambda_1$ can be written as $\ket{\boldsymbol{\gamma}}=\ket{\boldsymbol{\gamma_p }} \otimes \ket{\boldsymbol{\gamma}_s}$, where $\ket{\boldsymbol{\gamma_p}}=\int d\boldsymbol{p} \, \mathrm{F}(\boldsymbol{p})\ket{\boldsymbol{p}}$ and $\ket{\boldsymbol{\gamma}_s}=(\ket{+1}+\ket{-1})/\sqrt{2}$. Therefore, it is a separable state. This is not true anymore for the received state $\ket{\boldsymbol{\gamma '}}$ observed at $\lambda_2$, since momentum and helicity states now appear entangled.

We are able to provide a reference state where no Wigner phase is acquired. To do this we use wavepackets of the Bell-type $(\ket{+_p-_p}\pm\ket{-_p+_p})/\sqrt{2}$ instead of the simple superposition of $\ket{\boldsymbol{\gamma}_s}$ states. Indeed, in such a Bell-type state each helicity eigenstate in the tensor products gives opposite phase contributions. Therefore, it is easy to check that the total state remains invariant under the rotation of the polarisation \cite{invariantQI,Bradler}.

On the contrary, one can increase the effect by using GHZ-like states of helicity $(\ket{+_p}^{\otimes m} \pm \ket{-_p}^{\otimes m})/\sqrt{2}$. In this case, the $\ket{+_p}^{\otimes m}$ component acquires $m$ times the individual Wigner phase $\Psi$, while the $\ket{-_p}^{\otimes m}$ component picks up a total phase of $-m \Psi$. The total relative phase acquired by these states is therefore $2m \Psi$. In other words, there is an increase of the effect by a factor $2m$. One can also use a product of $m$ qubit states, namely $(\ket{+}+\ket{-})^{\otimes m}$, which is easier to prepare in the lab.

We have defined all the mathematical formalisms required in the rest of the work. We proceed to the next section where we compute explicitly the Wigner phase \eqref{Wigner phase} between two static observers in the background spacetime around the Earth.

%---------------------------------------------------------------------------------------------------------------------------------------%
\section{Polarisation rotation and Wigner phase in Schwarzschild spacetime}\label{polarisationrotation}
%---------------------------------------------------------------------------------------------------------------------------------------%
In this section, we proceed and compute the Wigner rotation of the polarisation vector of a light ray sent and received by static observers at different heights in a gravitational field. We then compute the associated Wigner phase acquired by the quantum state of the photons. 

%---------------------------------------------------------------------------------------------------------------------------------------%
\subsection{Polarisation rotation for static observers}\label{Earthobserver_rot}
%---------------------------------------------------------------------------------------------------------------------------------------%
Our goal requires us to construct the tetrad field for the chosen observers.
The tetrad field associated to the family of static observers along the light ray's null geodesic has $\boldsymbol{e_{\hat{0}}}(\lambda)=\boldsymbol{v_E}\left(r(\lambda)\right)$ as its zeroth component, where $r(\lambda)$ stands for the radial coordinate of the photon at affine parameter $\lambda$. We then follow the adaptation procedure introduced above \eqref{adapting}, and we use the expression \eqref{null} for a general null vector, with $k^{\hat{0}}=\boldsymbol{k}\cdot\boldsymbol{e_{\hat{0}}} \, \eta^{\hat{0}\hat{0}}=-\boldsymbol{k}\cdot\boldsymbol{v_E}=E_p/\sqrt{f(r)}$. We obtain the expression of the third triad component $\boldsymbol{e_{\hat{3}}}$ which reads
\begin{align}\label{tetrad3rd}
\boldsymbol{e_{\hat{3}}} =&\,  \sqrt{f(r)} \Bigg( \epsilon_r \sqrt{1-f(r)\frac{l_{\phi}^2+\kappa}{r^2}} \, \partial_r \nonumber \\
&\,+ \frac{\epsilon_{\theta}}{r} \sqrt{\frac{\kappa-l_{\phi}^2 \cot^2\theta}{r^2}} \, \partial_{\theta} + \frac{l_{\phi}}{r^2 \sin^2\theta} \, \partial_{\phi}\Bigg).
\end{align}
The remaining triad components $\boldsymbol{e_{\hat{1}}}$ and $\boldsymbol{e_{\hat{2}}}$ are then obtained using the orthonormalisation relations, as explained in Section \ref{tetrad_formalism}. We give their cumbersome expressions in appendix \ref{gen_tetrad}. It should be clear from the form of these expressions that the computation of the Wigner rotation $\tilde{\Psi}$ for the general null vector \eqref{null} and for the general adapted frame would be very involved.

In order to simplify the computations, let us now consider light rays constrained to a plane with constant longitude $\phi$, with null vector \eqref{null_constphi}. This is a reasonable assumption in a spherically symmetric spacetime where the light's trajectories are confined to propagate on a plane. The third triad component \eqref{tetrad3rd} then reduces to $\boldsymbol{e_{\hat{3}}} = \sqrt{f(r)} \left( \epsilon_r \sqrt{1-f(r)\frac{\kappa}{r^2}} \, \partial_r  + \frac{\epsilon_{\theta}}{r} \sqrt{\frac{\kappa}{r^2}} \, \partial_{\theta}\right)$. We do not display the Wigner rotation we obtain since its very involved expression is not particularly enlightening; however, we note that it always vanishes for radial trajectories, namely with $\kappa=0$, irrespectivelly of the frame choice we make. This indicates that the coupling between the Schwarzschild mass and the helicity of the photons occur only when the later have some angular momentum. We then make a simplifying choice for the reference frame: let us give the value $B=\frac{1}{r}\sqrt{1-\frac{\kappa}{r^2}}$ to the free component of the tetrad in \eqref{tetrad_general}. This choice implies that the expression of the Wigner rotation vanishes in the flat spacetime limit $M\to0$. Finally, to uniquely fix the reference frame, we set $\eta_1=\eta_2=+1$. The gauge dependent tetrad components associated to this choice of reference frame are displayed in appendix \ref{cho_tetrad}. We now proceed to calculate the Wigner rotation using \eqref{connection}, \eqref{tetrad_choice1}, \eqref{tetrad_choice2} and \eqref{null_constphi}. Expanding up to lowest order in the dimensionless perturbative parameter $\epsilon=\sqrt{M/R_E}\ll1$, we find
\begin{align}\label{static_rot}
\tilde{\Psi}= - \frac{\epsilon_r}{r} \frac{3 \, r^2 - \kappa}{r^2 - \kappa} \sqrt{\frac{\kappa}{2 \, r^2} \frac{R_E}{r}} \, \epsilon + O \left( \epsilon^3 \right).
\end{align}
For a light ray trajectory with a vanishing constant $\kappa$, there is no Wigner rotation as measured in the chosen frame. Such trajectories correspond to radial light rays, where both the polar angle $\theta$ and the longitude angle $\phi$ remain constant along the null geodesic. Notice that the Wigner rotation \eqref{static_rot} is independent on the energy of the photons which is encoded in $E_p$. This feature is already known in the literature \cite{Ca.re} and it remains true for the Wigner phase that will be computed in the next subsection. Note that one can obtain a reference channel where no Wigner rotation occurs by making the choice $B=0$ or $B=\pm\frac{1}{r}$ instead of the choice we made above.

%---------------------------------------------------------------------------------------------------------------------------------------%
\subsection{Wigner phase between static observers}\label{Earthobserver_Wigner}
%---------------------------------------------------------------------------------------------------------------------------------------%
We can now proceed and compute the Wigner phase $\Psi$ acquired by the momentum-helicity quantum states of photons travelling through Schwarzschild spacetime, from $r(\lambda_1)=R_1$ to $r(\lambda_2)=R_2$. We employ $\Psi =\int_{R_1}^{R_2} \tilde{\Psi} \, \frac{dr}{u^r}$ and \eqref{static_rot}, and obtain to lowest order{\small
\begin{align}\label{static_phase}
\Psi =  \left( \sqrt{\frac{R_E}{R_2}} \sqrt{\frac{2\,\kappa}{R_2^2 - \kappa}}  -\sqrt{\frac{R_E}{R_1}} \sqrt{\frac{2\,\kappa}{R_1^2 - \kappa}} \right) \epsilon+ O \left( \epsilon^3 \right).
\end{align} }
We emphasize that, for the static observers and for the frame choice we have made, there would be no Wigner phase for a light ray with a purely radial trajectory, i.e., $\Psi=0$ for $l_{\phi}=\kappa=0$. Note that $\kappa$ can only take a value up to $\kappa_{\textrm{max}} = \textrm{min} \Big\{ \frac{R_1^2}{1-\frac{2M}{R_1}} , \frac{R_2^2}{1-\frac{2M}{R_2}} \Big\}$. This means that $\kappa$ never reaches $R_1^2$ or $R_2^2$, therefore \eqref{static_phase} is always finite.

In the next section we apply our result \eqref{static_phase} to schemes where quantum optical signals are exchanged between laboratories on Earth using the aid of reflecting mirrors placed on satellites.

%---------------------------------------------------------------------------------------------------------------------------------------%
\section{Polarised photonic signals in Earth-satellite schemes}\label{schemes}
%---------------------------------------------------------------------------------------------------------------------------------------%
The schemes described in this section consist of a station placed on Earth that emits polarised photonic pulses of light to a satellite, which in turn reflects these signals to another satellite or to another laboratory on Earth. The signal is then communicated back to the emitter's laboratory in order for the trajectory of the pulse to form a closed path. We are interested in those schemes where light travels closed paths since it is possible to compare polarisation states in a ``natural'' way only when the associated light rays have parallel null vectors \cite{Pa.Ta.We}. This is a consequence of the fact that the polarisation vector is always defined by its orthogonality to the light's null vector, i.e., $\boldsymbol{k}\cdot\boldsymbol{\psi}=0$. Therefore, to talk about polarisation without specifying the corresponding null vector (or momentum) is meaningless. One can thus only compare polarisations associated to parallel null vectors.

In curved spacetime this comparison is not easy to achieve in general, unless in stationary spacetime backgrounds (like in the present study), where the light ray is bound to back to the same space location it was emitted from, and reflected to acquire the same initial propagation direction. To align the received light ray with the emitted one, it suffices to have a mirror at the reception location, oriented in such a way that the light ray would take the same propagation direction that it had when it was emitted. Placing the detection device very close to the emitting source after this last reflection would thus enable to measure photons which null vectors are parallel to the emitted ones.

%---------------------------------------------------------------------------------------------------------------------------------------%
\subsection{Earth-satellite(s)-Earth schemes}\label{the_schemes}
%---------------------------------------------------------------------------------------------------------------------------------------%

We consider here setups where the light rays follow trajectories with a constant longitude, for which we have obtained the explicit expression of the Wigner phase \eqref{static_phase}. This requires that the laboratories on Earth and the satellite's reflections events are all placed at the same longitude $\phi_0$. The satellites themselves, however, are not constrained to the $\phi=$ const. plane and can have an arbitrary motion as long as they are located at longitude $\phi_0$ when the reflection event occurs. 
%\footnote{Note that in practice, due to the Earth's rotation, the signals reflected by the satellite would need to be received at a slightly different longitudinal spacetime coordinates in order to be at the same longitude on Earth's surface. However, we ignore this additional effect here since we assume the Earth to be static since we are using the Schwarzschild metric to model the spacetime around the Earth. We leave it to further work to assess the practical implications of these additional effects.}
\begin{figure}[h!]
\includegraphics[width=8.5cm]{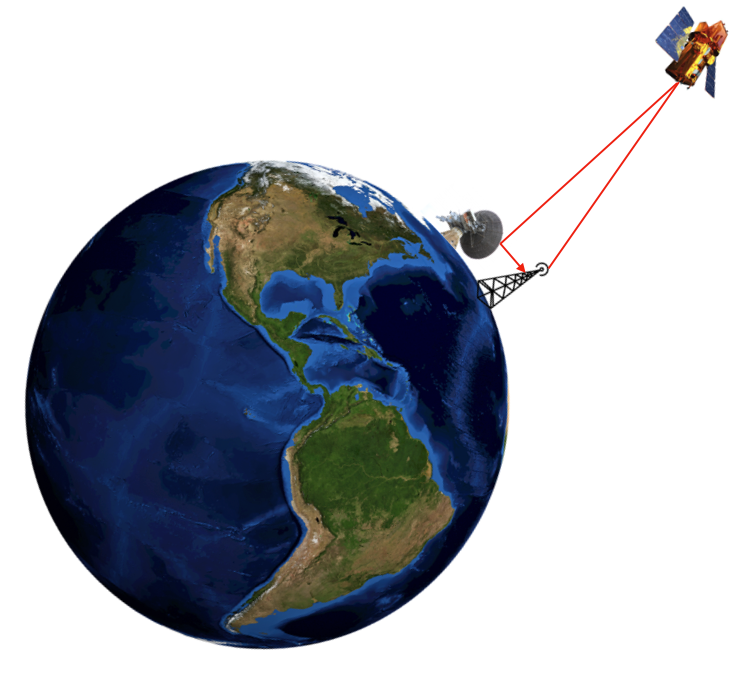}
\caption{
First scheme. Quantum optical signals (red line) are emitted from a station on Earth to a satellite which reflects them back to a second laboratory on Earth. From here, they are reflected back to the initial laboratory in order to close the path into a loop. The figure is depicted at a longitude $\phi=$ const. plane.}\label{figscheme1}
\end{figure}

The first scheme consists of two neighbouring laboratories placed on Earth at the same longitude, but at different polar angles, see Figure \ref{figscheme1}. The optical signals are emitted from one station and propagate to a satellite, they are reflected by the satellite to the second laboratory (in the same fashion as in recent satellite experiments \cite{Villoresi1,Villoresi2}), which in turn sends them back to the first laboratory. Note that if we were to simply reflect the light rays back from the satellite to the first laboratory there would be no Wigner rotation.

\begin{figure}[h!]
\includegraphics[width=8.5cm]{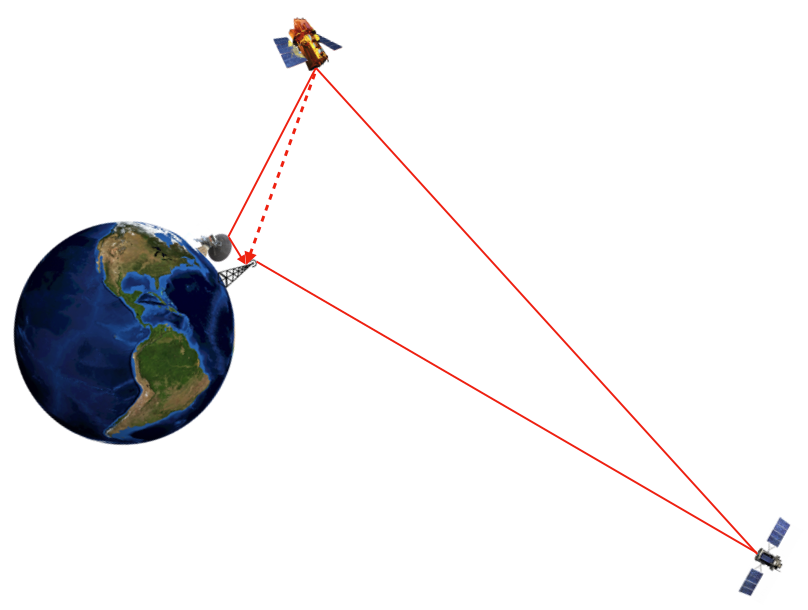}
\caption{
Second scheme. The light ray (the red line) is sent from Earth to a first satellite that reflects it to a second satellite, which in turn reflects it back to Earth either directly to the emitting station (dotted line), or to another laboratory (thick line). In the latter case, the signal is finally sent back to the initial laboratory in order to form a closed path. The figure is depicted at a longitude $\phi=$ const. plane for consistency with our calculations.}\label{figscheme2}
\end{figure}

Another scheme we will consider consists of a station on Earth emitting light pulses towards a satellite, which then reflects them towards a second satellite, in turn reflecting the signals back to the emitter's laboratory, see Figure \ref{figscheme2}. Alternatively, the downlink from the second satellite to Earth could also reach a second station that would then send the photons back to the emitter's laboratory, as shown in Figure \ref{figscheme2}. Since the light ray's radial coordinate remains approximately constant along the null geodesic in an extra small Earth-to-Earth segment, there is no Wigner phase contribution \eqref{static_phase} that arises from it, and the two variants for this scheme are equivalent.

%---------------------------------------------------------------------------------------------------------------------------------------%
\subsection{Wigner phase}\label{schemes_phase}
%---------------------------------------------------------------------------------------------------------------------------------------%
We have described the physical implementation of the two schemes of interest. We proceed and give the explicit expression for the total Wigner phase accrued by the photons along the closed paths designed above.

%---------------------------------------------------------------------------------------------------------------------------------------%
\subsubsection{Wigner phase: one satellite}
%---------------------------------------------------------------------------------------------------------------------------------------%
For the first scheme, we use \eqref{static_phase} for each different geodesic segment of the closed trajectory, and find the full expression for the Wigner phase $\Psi_1$, which reads
\begin{align}\label{phase_scheme1_gen}
\Psi_1 =&\Bigg[ \sqrt{\frac{R_E}{R_s}} \Delta K_s(\kappa,\kappa')+\Delta K_E(\kappa',\kappa)\Bigg] \epsilon +O(\epsilon^3),
\end{align}
where we have denoted the orbital radius of the satellite by $R_s$ and we have defined $K_X(y):=\sqrt{\frac{2\,y}{R_X^2-y}}$ and $\Delta K_X(y,z):=K_X(y)-K_X(z)$.

The Wigner phase contribution for the last segment of the light's propagation vanishes since its endpoints are at the same altitude, see \eqref{static_phase}. 
This is also the case if the receiver's laboratory is at the same location as the emitter's one, since then $\kappa'=\kappa$ and $\Psi_1=0$. This is the reason why, in the first scenario, we need to have the emitter and the receiver at different locations, i.e., at different polar angles. 

%---------------------------------------------------------------------------------------------------------------------------------------%
\subsubsection{Wigner phase: two satellites}
%---------------------------------------------------------------------------------------------------------------------------------------%
We now turn our attention to the second scheme. We repeat the procedure above and obtain the total Wigner phase $\Psi_2$, which reads
\begin{align}\label{phase_scheme2_gen}
\Psi_2 =&\Bigg[ \sqrt{\frac{R_E}{R_{s_2}}} \Delta K_{s_2}(\kappa',\kappa'')+ \sqrt{\frac{R_E}{R_{s_1}}} \Delta K_{s_1}(\kappa,\kappa') \nonumber \\
& +  \Delta K_E(\kappa'',\kappa)  \Bigg] \epsilon + O(\epsilon^3).
\end{align}
The radial coordinate of the first and the second satellite used to reflect the light ray are denoted by $R_{s_1}$ and $R_{s_2}$.

%---------------------------------------------------------------------------------------------------------------------------------------%
\subsubsection{Wigner phase: an explicit example}
%---------------------------------------------------------------------------------------------------------------------------------------%
We can obtain more physical insights by considering trajectories with $\kappa \ll R_{\textrm{min}}^2$, where $R_{\textrm{min}}=\textrm{min}\{R_1,R_2\}$, which amounts to consider almost radial light rays. In this constrained case, we have:
\begin{align}\label{Carter_cst_approx}
\kappa \approx \frac{R_1^2 \, R_2^2}{(R_1-R_2)^2}\Delta\theta^2.
\end{align}
This approximate expression is valid provided that $|\Delta\theta| \ll \frac{|R_1-R_2|}{R_{\textrm{max}}} =: \Delta\theta_c$, with $R_{\textrm{max}}=\textrm{max}\{R_1,R_2\}$. One can use \eqref{Carter_cst_approx} in scenarios such as when a light ray is exchanged between a laboratory close to the equator and a geostationary satellite. We can then expand the expression of the phase \eqref{phase_scheme1_gen} for $\kappa,\kappa' \ll R_E^2$:
\begin{align}\label{phase_scheme1_exp}
\Psi_1 \approx&\, \sqrt{2} (|\Delta\theta'|-|\Delta\theta|) \frac{1 - \left( \frac{R_E}{R_s} \right)^{\frac{3}{2}} }{1-\frac{R_E}{R_s}} \, \epsilon +O(\epsilon^3).
\end{align}
From this simple explicit formula, we see that when $|\Delta\theta'|=|\Delta\theta|$ there is no Wigner phase for this scheme. This feature is due to the symmetry of the spacetime, and it is simply a more explicit way of stating that \eqref{phase_scheme1_gen} vanishes for $\kappa=\kappa'$.

%---------------------------------------------------------------------------------------------------------------------------------------%
\subsection{Changes due to reflections on mirrors}\label{reflections}
%---------------------------------------------------------------------------------------------------------------------------------------%
Before proceeding to our main result, we review the particular effects induced by the reflections of the pulses by the moving satellites. In the schemes described in \ref{the_schemes}, the light rays are reflected by the satellite(s), and then by the mirrors of a static observer on Earth prior to their measurement since we need to align the momenta of the received photons with those of the emitted ones. This last reflection doesn't change the energy of the received photons as seen by the static observer that will perform the measurement, because the mirror is also static and located at the same distance $R_E$. Its only effects is to change the directional parameters $\epsilon_r$, $\epsilon_{\theta}$ and $\kappa$ of the null vector \eqref{null_constphi} back to the initial values.

However, the orbiting satellites \textit{do} impart some of their energy to the photons during the reflection events, which changes the value of the constant $E_p$ for the new geodesic segments that follow these reflection events.
The total state of the photon is affected by these events, and in particular the momentum states are. However, the Wigner phases are energy independent \cite{Ca.re}, therefore the helicity-part of the quantum states is not affected.

Finally, the reflections of light rays by the mirrors change the ray's direction, and therefore the polarisation vector as well. This will in principle induce extra Wigner phases, although we can safely ignore these effect in this work. This is a consequence of the fact that the relativistic corrections on the polarisation vector due to reflections on the mirrors are of one order higher in the perturbative expansion than the polarisation rotation induced by the propagation in the curved spacetime~\cite{PPN}.

%---------------------------------------------------------------------------------------------------------------------------------------%
\subsection{Wigner phase: existence in Schwarzschild}
%---------------------------------------------------------------------------------------------------------------------------------------%
We have shown that both of the total phases \eqref{phase_scheme1_gen} and \eqref{phase_scheme2_gen} are in general nonvanishing in Schwarzschild spacetime, and this can be seen even more explicitly for the first phase \eqref{phase_scheme1_gen} in its perturbative expansion \eqref{phase_scheme1_exp}. This is our main result. This result contradicts the statement found in the literature that the total Wigner phase acquired along a closed path in Schwarzschild spacetime would always vanish regardless of the chosen gauge convention \cite{brodutch_letter,brodutch}. 
The latter statement was a consequence of the general claim made by the authors, namely, that the Wigner phase remains gauge invariant under changes of reference frames along closed trajectories. In Schwarzschild spacetime, if one works within the Newton gauge, there is no rotation of polarisation along arbitrary null geodesics. The authors concluded that this absence of effect would remain along closed paths for any choice of reference frame. Yet, we have just shown in our derivation that a different choice of frame does not necessarily yield a zero result along a closed trajectory. In our view, the claim in \cite{brodutch_letter,brodutch} holds only for closed paths in phase space, which is a much stronger restriction. The paths we consider are not closed in phase space, although they form a closed trajectory in position space.

Note that in \cite{PPN} the authors first say that the geodetic effects of gravity on polarisation rotation are zero along a closed trajectory. However, in the same work they also say that this is true for closed phase space trajectories.
In our view, the geodetic effects vanish for closed paths in phase space only. 
An example is that of photons following a circular geodesic on the photon sphere of a black hole at $r_{\text{ps}} = 3M_{\text{bh}}$. In this case, the path is closed in both position and momentum spaces. One can expect the absence of Wigner phases for such configurations in Schwarzschild spacetime.
However, if the path is closed by judiciously positioned mirrors, like in this work, then the claim made in \cite{brodutch_letter,brodutch} does not hold. This occurs because the mirrors change the light's null vector in a discontinuous way, which therefore opens the path in momentum space. The arguments made by the authors in those works can no longer be applied, and the gauge-invariance is lost even if the photons are measured at the same location by the same observer who initially emitted them. This supports our nonzero result.

%---------------------------------------------------------------------------------------------------------------------------------------%
\section{Conclusion}\label{conclusion}
%---------------------------------------------------------------------------------------------------------------------------------------%
We have considered pulses of polarised light that travel through the curved spacetime around the Earth. In order to describe the evolution of photonic helicity states in curved spacetime backgrounds, we implemented the Wigner theory for massless spin-1 particles. We derived the general adapted frame for static observers and then obtained the Wigner phase for photons travelling between such observers in Schwarzschild spacetime. We have specialised to closed paths, where light is emitted and reflected by different links to reach again the emitter, and we found that the helicity quantum states of the photons pick up a non-zero Wigner phase.

We have studied two setups that could be implemented with current satellite technologies and we gave the expressions for the Wigner phases aquired by the photons along the path loops. We also gave a more explicit formula for a simplified version of the first scheme, where the labs would be located close to the equator and the satellite in geostationary orbit. All the phases we have obtained are non-zero except for some particular symmetries of the setups, and their values are specific to the frame choice that we made. Our result questions the statement found in the literature that there is no Wigner phase accrued by photons propagating along a closed path in Schwarzschild spacetime. This finding seems to indicate that the present theory is incomplete for the description of quantum measurements of polarisation effects in simple realistic schemes involving satellites, such as the ones we have described in this work.

We believe that the effects we found occur because the paths we considered are closed thanks to well positioned mirrors. If the paths were closed because of initial conditions, namely if the photons were following a single null geodesic from emission to measurement, then the gauge-invariance claims made in \cite{brodutch_letter,brodutch} should still hold. A detailed study of the Wigner phase acquired by photons at the reflection events seems indicated. We leave this investigation to further work.

%----------------------------------------------------------------------------------------------------------------------------------------------------------------------------------------------------------------------------%
\section*{Acknowledgments}
%----------------------------------------------------------------------------------------------------------------------------------------------------------------------------------------------------------------------------%
We thank Lars Andersson, Daniel Hartley, J\'er\'emie Joudioux, Jackson Levi Said, Luca Mazzarella, Marius Oancea and Daniel Oi for useful comments and discussions. This work was in part supported by the Anglo-Austrian Society. DEB acknowledges the COST Action CA15117 and CA15220 for partial support. JL was supported in part by Science and Technology Facilities Council (Theory Consolidated Grant ST/P000703/1). JK would like to thank the university of Vienna, Austria, for hospitality.

\appendix

%---------------------------------------------------------------------------------------------------------------------------------------%
\section{Tetrad}\label{tetrad_app}
%---------------------------------------------------------------------------------------------------------------------------------------%

%---------------------------------------------------------------------------------------------------------------------------------------%
\subsection{General tetrad}\label{gen_tetrad}
%---------------------------------------------------------------------------------------------------------------------------------------%
We give here the general first and second triad components of a static observer \eqref{static}, adapted to an arbitrary null vector \eqref{null}. Using the expressions of the zeroth $\boldsymbol{e}_{\hat{0}}=\boldsymbol{v_E}$ and third \eqref{tetrad3rd} tetrad components, together with the orthonormalisation relations, we have $e_{\hat{1}}^t= e_{\hat{2}}^t= 0$ and
\begin{widetext}
\begin{align}\label{tetrad_general}
e_{\hat{1}}^r=&\, - \eta_1 \, \epsilon_r \, (r-2M) \sin\theta \frac{r^4 l_{\phi}  B - \eta_2 \, \epsilon_{\theta} \sqrt{r^3 - (r-2M)(\kappa+l_{\phi}^2)} \sqrt{\kappa-l_{\phi}^2 \cot^2\theta} \sqrt{Q - r^5 B^2 \sin^2 \theta}}{r^2 \, Q}, \\
e_{\hat{1}}^{\theta}=&\, -\eta_1 \, \eta_2 \frac{\sqrt{Q - r^5 B^2 \sin^2 \theta}}{r^{\frac{5}{2}} \sin\theta}, \\
e_{\hat{1}}^{\phi}=&\, \eta_1 \frac{r^4 B \sin^2\theta \sqrt{r^3 - (r-2M)(\kappa+l_{\phi}^2)} + \eta_2 \, \epsilon_{\theta} \, l_{\phi} \, (r-2M) \sqrt{\kappa-l_{\phi}^2 \cot^2\theta} \sqrt{Q - r^5 B^2 \sin^2 \theta}}{r^{\frac{5}{2}} \sin\theta \, Q}, \\
e_{\hat{2}}^r=&\, - \epsilon_r \, (r-2M) \frac{\epsilon_{\theta} \, r B \sin^2\theta \sqrt{r^3 - (r-2M)(\kappa+l_{\phi}^2)} \sqrt{\kappa-l_{\phi}^2 \cot^2\theta} + \eta_2 \, l_{\phi} \sqrt{Q - r^5 B^2 \sin^2 \theta} }{\sqrt{r} \, Q}, \\
e_{\hat{2}}^{\phi}=&\, \frac{-\epsilon_{\theta} \, r B \, l_{\phi} \, (r-2M) \sqrt{\kappa - l_{\phi}^2 \cot^2\theta} + \eta_2 \sqrt{r^3 - (r-2M)(\kappa+l_{\phi}^2)} \sqrt{Q - r^5 B^2 \sin^2 \theta} }{r \, Q},
\end{align}
\end{widetext}
with $Q:=\left(r^3 - (r-2M)(\kappa+l_{\phi}^2)\right) \sin^2\theta+ l_{\phi}^2 \, (r-2M)$ defined for compactness. The signs $\eta_1=\pm1$ and $\eta_2=\pm1$, and the tetrad component $B=e_{\hat{2}}^{\theta}$, are arbitrary. The latter quantity is however constrained to be of dimension inverse of a length. These three parameters encode all the gauge from the freedom of choice of our adapted frame.

%---------------------------------------------------------------------------------------------------------------------------------------%
\subsection{Chosen tetrad}\label{cho_tetrad}
%---------------------------------------------------------------------------------------------------------------------------------------%
In \ref{Earthobserver_rot} we set $l_{\phi}=0$ and we fixed the gauge parameters as $B=\frac{1}{r} \sqrt{1 - \frac{\kappa}{r^2}}$ and $\eta_1=\eta_2=+1$ in order to simplify the expression of the Wigner rotation. The gauge dependent tetrad components then simplify to
\begin{widetext}
\begin{align}\label{tetrad_choice1}
\boldsymbol{e}_{\hat{1}}=&\,  \epsilon_r \, \epsilon_{\theta} \, \frac{\kappa}{r^2} \sqrt{\frac{2M}{r}} \frac{1-\frac{2M}{r}}{\sqrt{1 -  \left(1-\frac{2M}{r}\right) \frac{\kappa}{r^2}}} \, \partial_r 
- \frac{1}{r} \sqrt{\frac{2 M \kappa}{r^3}} \, \partial_{\theta} 
+ \frac{\csc{\theta}}{r} \frac{\sqrt{1 - \frac{\kappa}{r^2}}}{\sqrt{1 - \left(1-\frac{2M}{r}\right) \frac{\kappa}{r^2}}} \, \partial_{\phi}, \\ \label{tetrad_choice2}
\boldsymbol{e}_{\hat{2}}=&\, - \epsilon_r \, \epsilon_{\theta} \sqrt{\frac{\kappa}{r^2}} \left(1-\frac{2M}{r}\right)\frac{\sqrt{1-\frac{\kappa}{r^2}}}{\sqrt{1 - \left(1-\frac{2M}{r} \right)  \frac{\kappa}{r^2}}} \, \partial_r 
+ \frac{1}{r} \sqrt{1 - \frac{\kappa}{r^2}} \, \partial_{\theta} 
+ \frac{\csc\theta}{r} \frac{\sqrt{\frac{2M\kappa}{r^3}}}{\sqrt{1 - \left(1-\frac{2M}{r}\right)\frac{\kappa}{r^2}}} \, \partial_{\phi}.
\end{align}
\end{widetext}

\normalem
\bibliographystyle{apsrev4-1}
\bibliography{helicity}

\end{document}